\documentclass[10pt]{llncs}
\usepackage{amssymb, amsmath, graphicx,subfigure, slashbox,hyperref,amssymb, booktabs, tabularx,threeparttable}

\newcommand{\tabincell}[2]{\begin{tabular}{@{}#1@{}}#2\end{tabular}}
\begin{document}
%
\frontmatter

\title{Classroom Video Assessment and Retrieval via Multiple Instance Learning}

\author{Qifeng Qiao \and Peter A. Beling}
\institute{Department of Systems and Information Engineering
\\ University of Virginia, USA
\\ \email{qq2r, pb3a@virginia.edu}}

\maketitle

\begin{abstract}
We propose a multiple instance learning approach to  content-based retrieval of classroom video for the purpose of supporting human assessing the learning environment.   The key element of our approach is a mapping between the semantic concepts of the assessment system and features of the video that can be measured using techniques from the fields of computer vision and speech analysis.  We report on a formative experiment in content-based video retrieval involving trained experts in the Classroom Assessment Scoring System, a widely used framework for assessment and improvement of learning environments.   The results of this experiment suggest that our approach has potential application to productivity enhancement in assessment and to broader retrieval tasks.
\end{abstract}

\section{Introduction}
\label{section:intro}
Classroom assessment is a topic of increasing interest among education practitioners, researchers, and policy makers.  Recent years have seen a number of  observation and assessment protocols developed, fielded, and tested as part of large-scale effectiveness experiments.   The Measure of Effective Teaching (MET) project, for example, is designed to help educators and policy makers identify and support good teaching by improving the quality of information about teacher practice.  MET has used approximately 500 assessment experts, known as coders, to rate more than 23,000 hours of videotaped lessons using standard classroom observation protocols.  Recent years also have seen advances in the fields of computer vision and machine learning,  to the point where it is reasonable to consider a role  in the classroom assessment process for automatic interpretation of video, audio, and other sensor information.  In the near-term, this role is likely to be one of supporting, rather than supplanting, human coders by providing filtering or pre-screening services to distill large volumes of video down to those portions that are likely to be most productive or informative for assessment.   

We assert that content-based video retrieval is a core technical problem for the development of filtering schemes.  The aim in content-based retrieval is to use training interaction with a human user to gain an understanding of the media content that is of interest to the user.  
Content-based image retrieval has been widely studied, and recently there has been some extension of this work to video, with focus on entertainment media like television programs and feature films. Classroom videos have a number of idiosyncratic properties that present both challenges and opportunities in retrieval.   Difficulties in interpretation arise from the complicated and dynamic nature of classroom events, occlusion among students, and pragmatic aspects of human communication.  On the other hand, the structured environment of a classroom means that, within the context of a particular assessment methodology, it may be possible to decompose dynamic events into a set of simpler components that are amenable to machine measurement.   

In this paper, we propose the Classroom Evaluation and Video Retrieval (CLEVER) system, which is a multiple instance learning (MIL) approach to content-based retrieval of classroom video for the purpose of supporting human assessing the learning environment.   The learning aspects of CLEVER are similar to MIL and other approaches that have been used for content-based image and video retrieval (cf. \cite{qiao2009,uijlings2010,fan2004}), but differ in that instances and the feature space are defined in ways that exploit the structure of classroom learning and the nature of the assessment system.  The key element in CLEVER is a mapping between the semantic concepts of the assessment system and features of the video that can be measured using techniques from the fields of computer vision and speech analysis.   We work with a single assessment methodology, the Classroom Assessment Scoring System (CLASS).   
CLASS  is a theoretically-driven and empirically-supported conceptualization of classroom interactions \cite{pianta2008} in which trained coders produce assessment scores on the basis of observation of the classroom, either in person or from a video recording or broadcast.  The framework encompasses a consultive process in which teachers used annotated video, produced by the coders using a structured process, as the basis for a self-improvement effort \cite{pianta2007}.  CLASS has been widely adopted, earning places in both Head Start and MET assessment projects.  The CLASS methodology centers on observation of teacher and student actions and interactions, a behavioral orientation that tends to align well with machine interpretation of video, particularly in comparison with assessment approaches that focus on instructional content.

The remainder of the paper is organized as follows: In Section \ref{section:model}, we present a mapping between the structure of CLASS and concepts that have associated measurements created through automated processing of video and audio. We also describe the multiple instance framework that is the basis for our learning method. In Section \ref{section:experiments}, we  report on the use of CLEVER in a formative experiment in content-based video retrieval involving a group of expert CLASS coders. Finally, in Section \ref{section:discussion}, we offer conclusions and suggestions for future research.

\section{Video Understanding in CLASS}
\label{section:model}

The CLASS framework is a theoretically-driven and empirically validated conceptualization of classroom interactions  \cite{pianta2005a,pianta2007,pianta2008}.    CLASS embodies a latent structure for organizing classroom activity in three domains: {\em emotional support}, {\em classroom organization}, and {\em instructional support}. Each domain is composed of several dimensions defined semantically and scored quantitatively \cite{pianta2008}.  To take one example, the dimension {\rm productivity}, within {\em classroom organization} can be classified into three levels: low with a score of 1 or 2, medium with a score of 3, 4 or 5 and high with a score of 6 or 7.  The high level would be assigned to a classroom in which students are oriented, with respect to expectations and tasks, and transitions from one activity to another happen quickly and efficiently.  CLASS coders rely on their judgment and reasoning intelligence to assign scores.

\begin{figure}[!t]
	\centering
	  \includegraphics[width = 3.5in]{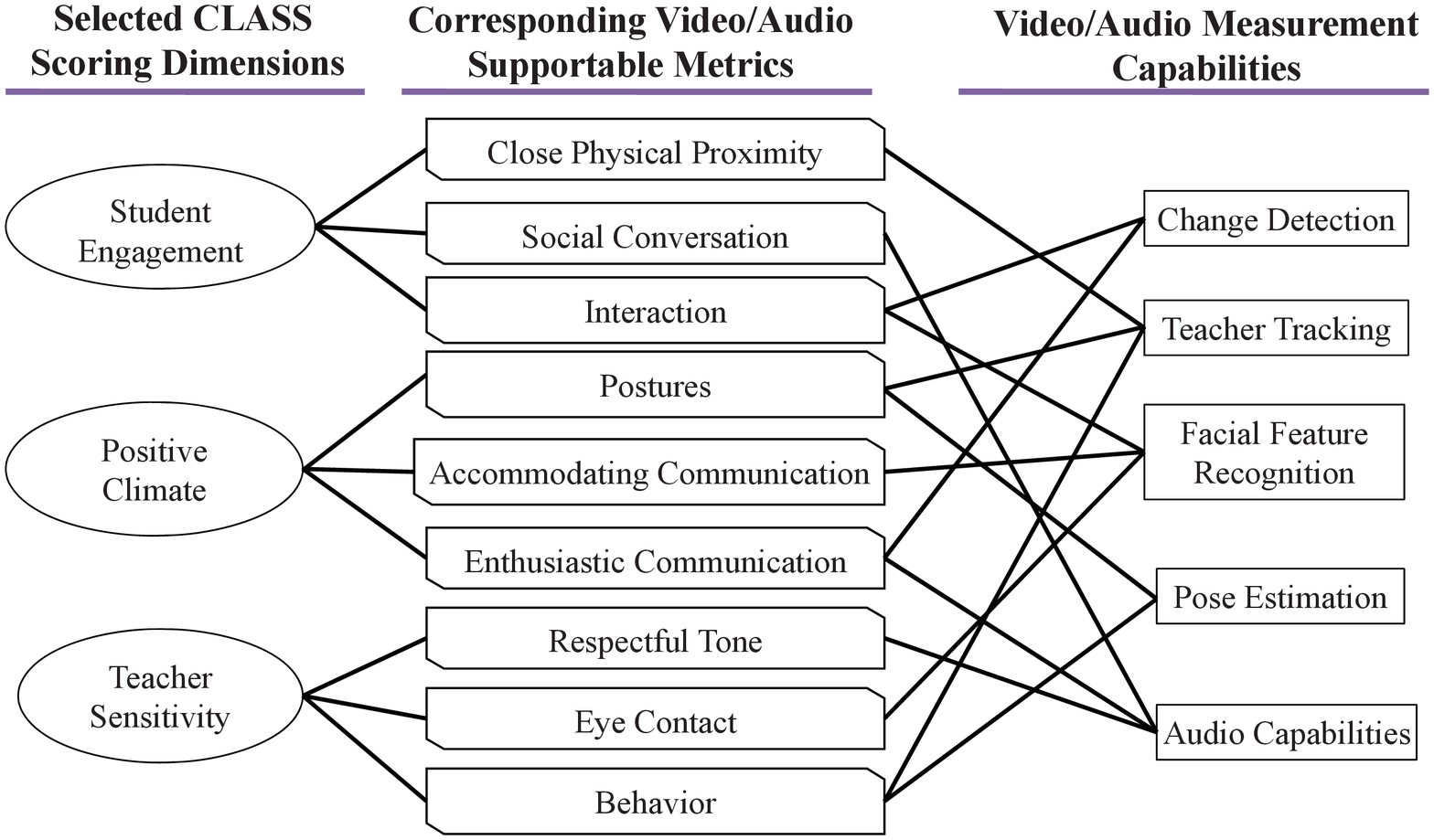}
	  \caption{Example of semantic gap bridging between CLASS and automatic measurement}
	  \label{figure:CLASSmap}
\end{figure}

\subsection{Feature Extraction}
\label{section:feature}

An ideal video retrieval system would allow one to query on high-level concepts, often called {\em semantic concepts}. As an example, one might like to ask a retrieval system for all classroom videos in which the teacher appears to be frustrated with student progress or those that present a high level of energy on the part of the students.  Automated retrieval systems, however, must work with much lower level concepts, such as pixel intensity and pixel change or sound frequency, that can be measured from video and audio using algorithms.  The principal challenge in video retrieval is to bridge the gap between semantic concepts and {\em measurable concepts}, which are the features we can handle using automated video interpretation.  In our case, the scoring dimensions of CLASS are the relevant semantic concepts.  As they relate to classroom assessment, we call these {\em semantic assessment concepts}.  A good semantic-sensitive video content representation framework emphasizes features that are more capable of representing the semantic assessment concepts and avoids performing uncertain feature extraction. For example, the semantic assessment concepts of instructional aiding materials, lecture presentation, and student engagement are implicitly related to visual analyses, including the detection of moving objects, high luminosity regions, human faces or skin, and blocks of changing pixels, as well as audio analyses, such as detection of individual and dialog speech. 

As illustrated in Fig.\ref{figure:CLASSmap}, we propose bridging the gap between semantic assessment concepts and measurable concepts in two steps, first linking semantic assessment concepts with video/audio metrics from CLASS dimensions, and then relating the video/audio metrics with feature variables that can be extracted by available automatic measurement techniques.  Many video processing techniques we need, such as topical detection, synchronization, summarization and editing, have been addressed for content analysis of classroom videos \cite{wang2007}. These tasks depend on static analysis of image features, e.g. detection of the slides using color background detection \cite{syeda2000}, key-frame detection using similarity measurement  and scene-break detection using image differences and color histograms \cite{ju1998}.  Making use of the relationship between CLASS and video/audio measurement capability, we characterize classroom videos using the attributes in Table \ref{table:feature}.  The measurement of high-level attribute requires combination of multiple feature extraction techniques. For example, group discussion events are found using the lower-level features of speech detection and motion intensity estimation.

\begin{table}[!t]
\caption{Video feature definition used to construct the feature vector for each video}
{\footnotesize \begin{tabularx}{\textwidth}{lX}
\toprule
\textbf{ Low-level Attribute} & \textbf{ Description}\\
\midrule
 Color Histogram & Global color represented in HSV space\\
\midrule
 Co-occurrence Texture & \tabincell{l}{ Global texture containing entropy, energy, and contrast.}\\
\midrule
 Motion Intensity & \tabincell{l}{  Average difference of pixel values.}\\
\midrule
 Teacher Position &  Teacher's position in the classroom.\\
\midrule
 Moving Velocity & \tabincell{l}{ Mean, maximum, and minimum velocity of detected movement.}\\
\bottomrule
\textbf{ High-level Attribute} & \textbf{ Description}\\
\midrule
 Salient Object & \tabincell{l}{ Image regions with homogeneous color or texture.}\\
\midrule
 Pose Orientation &  Teacher's orientation: toward students or toward blackboard. \\
\midrule
 Teacher Gesture & \tabincell{l}{ Detection and recognition from a predefined gesture set.} \\
\midrule
 Dynamic Event & \tabincell{l}{ Student presentations, group discussion.}  \\
\bottomrule
\noalign{\smallskip}
\textbf{ Audio Attribute} & \textbf{ Description}\\
\midrule
 Silence Detection &  Silence on the part of the teacher\\
\midrule
 Pitch &   Frequency of speech\\
\midrule
 Dialog Talking & \tabincell{l}{ Question and answer events.}\\
\bottomrule
\end{tabularx} }
\label{table:feature}
\end{table}

\subsection{Multiple Instance Structure and Learning}
\begin{figure}[!t]
	\vspace{-10pt}
	\centering
		\includegraphics[width = 3.6in]{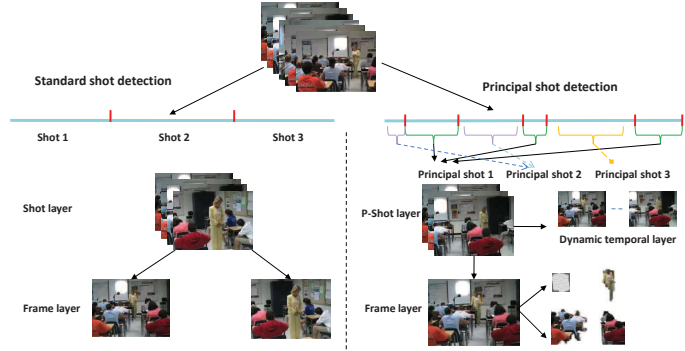}
	  \caption{Comparison of structures for video understanding. The left interprets the traditional structure and the right displays the principles of multiple instance structure.}
	  \label{figure:videoStruct}
\end{figure}

Most methods of shot boundary detection focus on segmenting the video clip at frames corresponding to transitions, either abrupt (cuts) or gradual (dissolves and fades). These shot detection techniques have limited application in our context because scene scenarios of classroom videos are relatively stationary and unvaried as measured by global low-level attributes, such as color histogram and textures. Moreover, in classroom video a measurable concept may appear in different temporal locations, implying the concept is represented by a set of small video sequences that are highly correlated. We propose a new framework that depends on an interpretation of the assessment protocol that varies according to individual perceptions.  In Fig. \ref{figure:videoStruct}, we show the traditional structure for shots, consisting of contiguous temporal regions, compared with our proposed \textsl{principal shot structure} in which shots are composed by aggregating segments from across the video that share a common semantic assessment concept.  This structure depends on both static and dynamic video patterns for video content representation and feature extraction. We expect such shot detection structure to enhance the quality of features since it gives rise to a hierarchical analysis of video content and an understanding of semantic objects and temporal events.  

We use MIL as the primary method for relating high-level concepts of interest to the user to measurable concepts. MIL is a variation of supervised learning in which there is ambiguity associated with labels \cite{dietterich1997}. Instead of receiving labels for each instance, the training set is composed of a number of {\em bags}, each of which is comprised of a set of instances.  In binary MIL, a bag is labeled positive if it contains at least one positive instance, and is labeled negative otherwise. Given labels for a set of training bags, the learning algorithm aims to discover the regions of the feature space associated with positive labels, with the particular goal of labeling individual bags and instances correctly.  A variety of algorithms have been developed for MIL, including \cite{qiao2009,Andrew2003,dietterich1997}.  

MIL has been successfully applied in the field of localized content based image retrieval (LCBIR) \cite{qiao2009,rahmani2008}, where the goal is to rank images according to their similarity to training images that a user has labeled as being of interest.   In LCBIR, images are the bags and contiguous blocks of pixels are the instances. In our application, video clips are the bags and principal shots are the instances.  We construct principal shots by first segmenting each video clip into micro clips (e.g. a segment of 10 seconds length). We then use adaptive k-means clustering  to group similar micro clips, with each group forming a principal shot. The general learning process includes: measurement and feature extractions, video segmentations, clustering of micro clips, feature aggregation for principal shots, MIL, and calibration with ground truth data.
 

\section{Experimentation with Human Evaluators}
\label{section:experiments}
\begin{table}[!t]
\footnotesize{
\caption{Performance Accuracy with respect to i-th subject (Si, $i=1,2,\cdots,10$)}
\label{table:acc}
		\begin{tabular*}{\textwidth}{@{\extracolsep{\fill}}ccccccccccc}
			\toprule
Video &   S1 &   S2 &   S3 &   S4 &   S5 & S6 &   S7 &   S8 &  S9 &  S10\\
\midrule
   Course A &      0.904 &      0.645 &      0.635 &      0.794 &      0.734  & 0.763 &      0.616 &      0.768 &        0.9 &      0.612\\
\midrule
   Course B &      0.898 &      0.524 &      0.652 &      0.636 &      0.622 & 0.652 &      0.668 &      0.678 &      0.994 &      0.686\\
\midrule
Mixed data &      0.901 &      0.585 &      0.644 &      0.715 &      0.678 & 0.707 &      0.642 &      0.723 &      0.947 &      0.649 \\
\bottomrule
\end{tabular*}}
\end{table}
As a formative experiment, we conducted an experiment with 10 expert CLASS coders. Coders asked to view 40 video clips, each three minutes in length.  Clips were taken from video recordings of two junior-level classes in Systems Engineering at the University of Virginia ({\em Course A} and {\em Course B}).  Coders were instructed to assign either a positive or a negative label to each clip, giving a positive label only if, in their individual judgment, the clip provided significant useful information for the purposes of CLASS assessment.  Coders were further instructed to evaluate each clip in isolation from the other clips, so that behavior or activity that had been seen before in the sequence was just as deserving of a positive label as when seen for the first time.  Coders were free to formulate their own interpretations of CLASS in relation to the labeling instructions.  Perhaps as a result, labels varied greatly across the subjects, with a Fleiss' kappa \cite{fleiss1971} value of 0.146 for Course A and 0.125 for Course B.   

For the basic experiment of learning labels, classification accuracy is defined as the proportion of the correctly predicted labels in the testing dataset.  We estimated classification accuracy on the basis of 100 replications, each with equal-sized, randomly chosen training and testing sets.  Results are shown in Table \ref{table:acc}.  The large variation in accuracy across subjects is likely a reflection of the variation in semantic concept reflected in the subjects' choices of labels.

\begin{figure}[!th]
	\centering
		\subfigure[]{\includegraphics[width =2.2in]{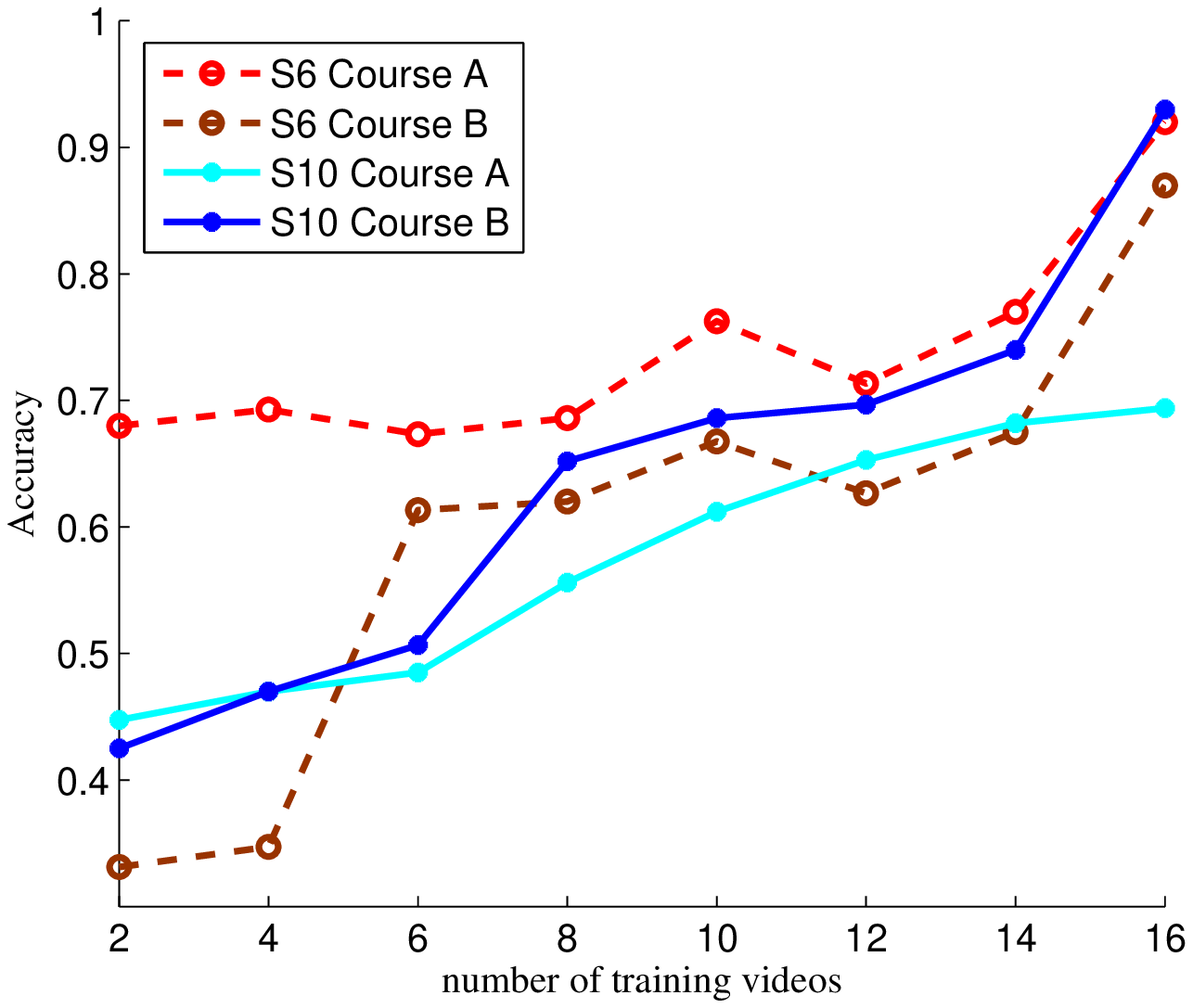}}
		\hfill
		\subfigure[]{\includegraphics[width =2.2in]{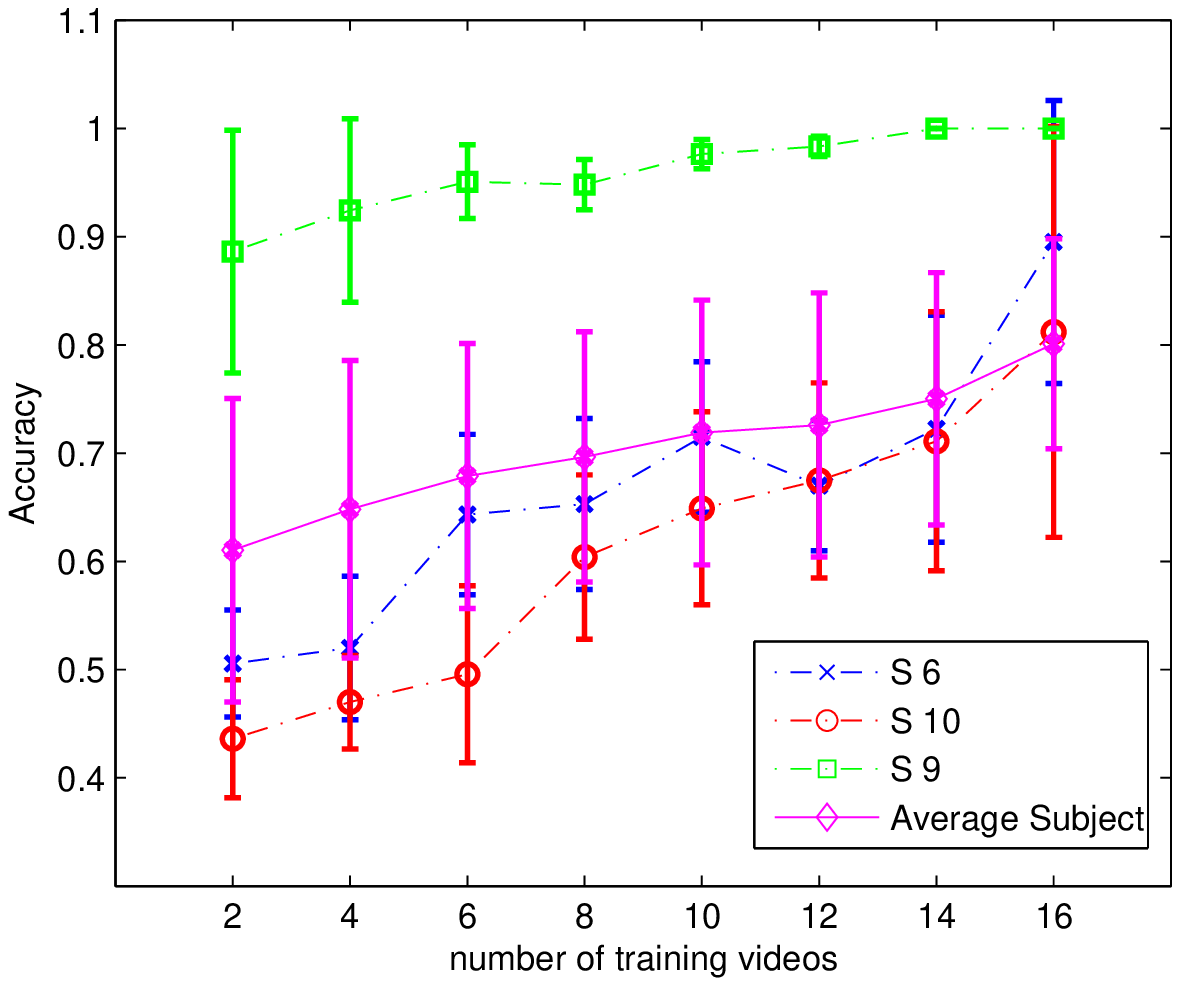}}
		\caption{Plots of Performance Accuracy. Fig (a) shows the mean classification accuracy 
		Fig (b) displays the mean accuracy and the standard deviation, where \textsl{average subject} represents the accuracy that is averaged across ten subjects.}
		\label{figure:learningcurve}
\end{figure}

To investigate consistency of predictive performance, we estimated classification accuracy as a function of training set size. Fig. \ref{figure:learningcurve} shows this relationship for three coders.  In these examples CLEVER performance on individual user is consistent, since accuracy is increasing in the number of training examples used. Average performance across the 10 subjects exhibits the same trend.  

\begin{figure}[!th]
	\centering
		\subfigure[Theoretic computation]{\includegraphics[width =2.2in]{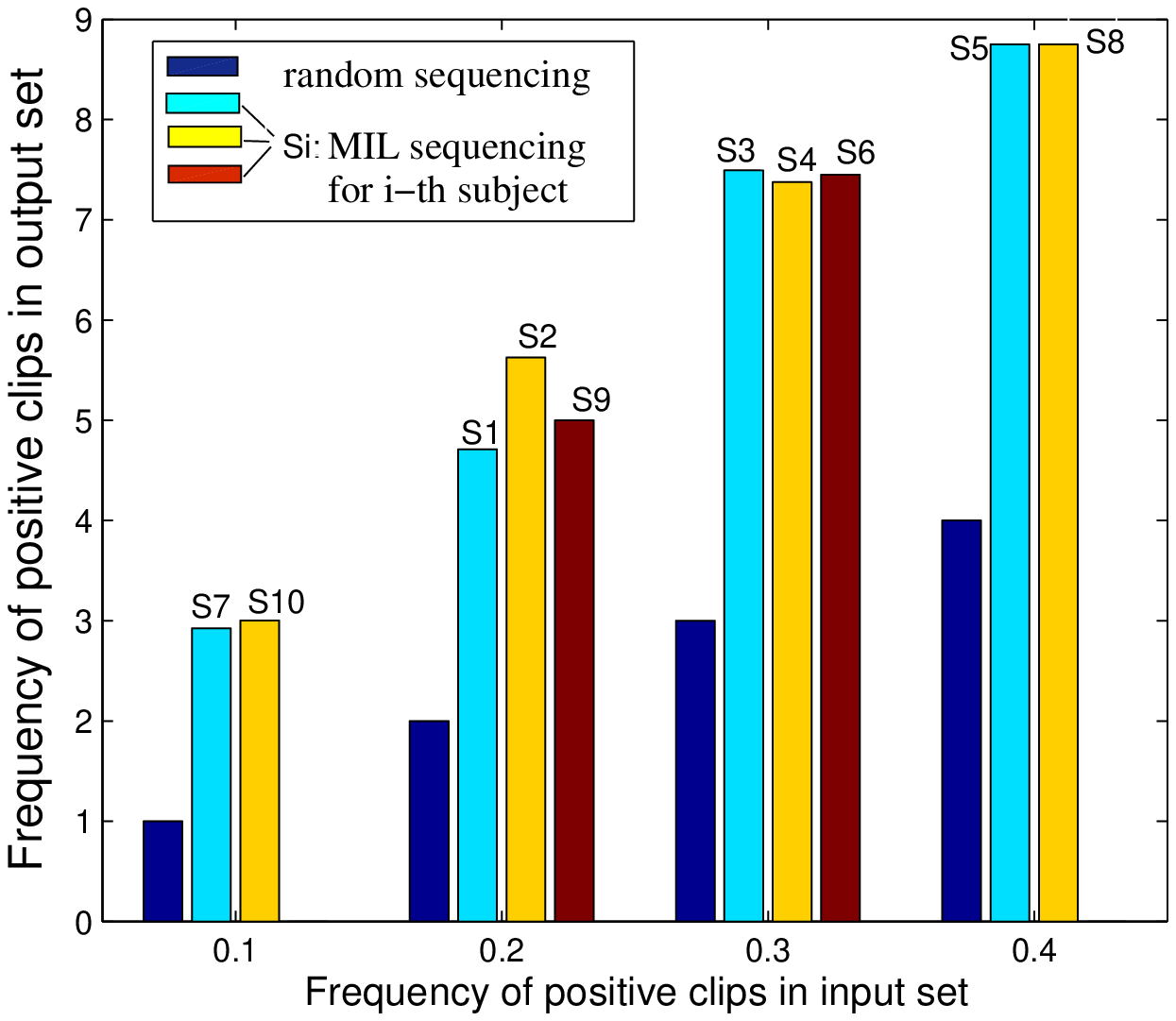}}
		\subfigure[Experimental simulations]{\includegraphics[width =2.2in]{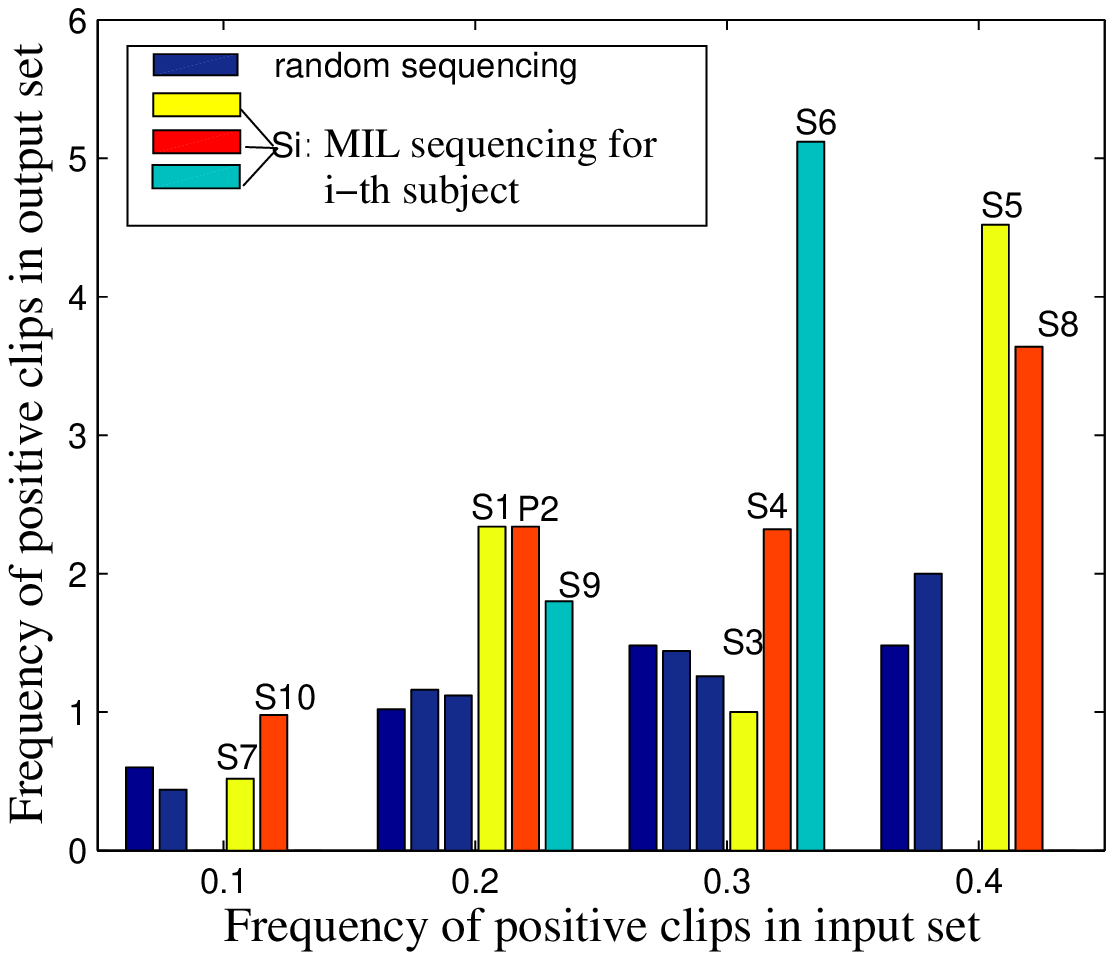}}
		\caption{Plots of productivity. The productivity bars for ten subjects are shown in groups with regard to the frequency of positive clips in input set. }
		\label{figure:productivity}
\end{figure}

To investigate potential filtering roles for CLEVER, we used the label data from the coders in computational experiments on productivity.  The setting for these experiments is a hypothetical scenario in which a coder is viewing a sequence of video clips.   The machine learning task is to use labels from the first 10 minutes of viewing to reorder the remaining clips with the goal of maximizing the number of positive clips viewed during a 10-minute performance period.  Figure \ref{figure:productivity} (a) compares the expected number of positive clips viewed under a random order with that expected from a reordering done with an accuracy equal to the estimated true positive probability achieved by CLEVER label learning experiments described above.  The reordering outputs the predicted positive videos and we assume there is enough predicted positive videos for viewing in 10 minutes. Fig. \ref{figure:productivity} (b) shows the results for a similar computation that, instead of estimated accuracies, used  100 replications of  the simulation on a boosted testing data set containing 100 positive and 100 negative clips.

\section{Discussion and Future Work}
\label{section:discussion}
CLEVER fuses state-of-the-art machine learning algorithms with advanced assessment concepts from the education community.  The results of formative experiments on CLASS coders are encouraging.  Accuracy in label prediction is substantially greater than would be expected from random performance and, as our productivity experiments show, would be sufficient to support filters that would reduce human viewing load by a factor of 2 or more.  It is also worth noting that other users, such as teachers themselves, might benefit from the content-based retrieval capability of CLEVER as part of a self-improvement or reflective process.

The feature set that we used could be improved through the addition of  more audio characteristics.  Furthermore, the integration of video and audio techniques might be the key to extracting higher-level features that underscore interactions between teacher and students, which are known to be critically important elements of classroom assessment. 

\bibliographystyle{unsrt}
\bibliography{ClassviaMIL}

\end{document}